# The physics of business cycles and inflation

Hans G. Danielmeyer and Thomas Martinetz
Institut für Neuro- und Bioinformatik, Universität zu Lübeck, Germany
www.uni-luebeck.de

**Abstract**
We analyse four consecutive cycles observed in the USA for employment and inflation. They are driven by three oil price shocks and an intended interest rate shock. Non-linear coupling between the rate equations for consumer products as prey and consumers as predators provides the required instability, but its natural damping is too high for spontaneous cycles. Extending the Lotka-Volterra equations with a small term for collective anticipation yields a second analytic solution without damping. It predicts the base period, phase shifts, and the sensitivity to shocks for all six cyclic variables correctly.

   Business cycles are more or less regular fluctuations of production, employment, consumption, growth, money supply, and inflation on the time scale of up to 7 years. They have nothing to do with long-term technical change or medium-term recovery of the technical infrastructure from wartime destruction. Their cause and control are central problems of economic policy [1], but their analytic understanding is still limited to accepting the oscillator [2] and Lotka-Volterra [3] equations without damping, or denying the existence of cycles because there seems to be no propagator [4]. Times of negative growth are treated with estimated money injections. Times of positive growth are approximated with the "Phillips Curve" for the relative change in % between annual inflation and employment [1].

   Figure 1 shows the only existing "experiment" with 4 consecutive cycles. Its initial wing from 1961 to 1966 follows the Phillips Curve. Due to a temporary entente between the USA and the former USSR the first cycle was initially too flat and too long. Due to the Vietnam War it ended nearly rectangular.

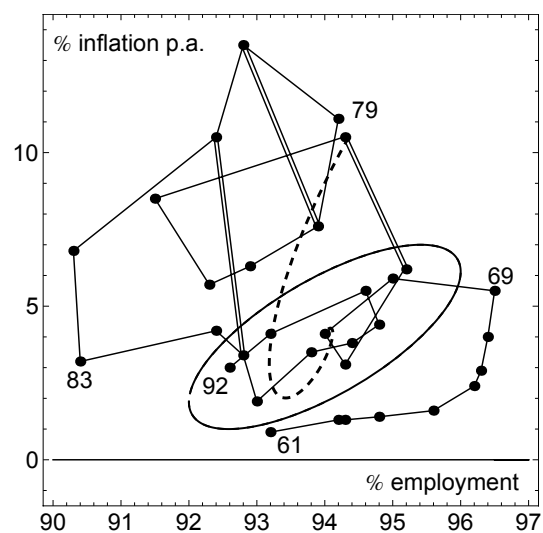

*Figure 1: Four business cycles of the USA. The data are taken from [1]. The doubled changes of the inflation rate are due to 3 crude oil price shocks. Natural damping prohibits cycles (dashed envelope). The added ellipse shows the normal range of cycles.*

   The original unemployment axis was reversed. The double lines show the increase of inflation due to the crude oil price shocks of the Jan Kippur Embargo and the Iranian Revolution, and its return to the normal range towards the end of the Cold War. The first price change happened in the calendar year 1974, the second in two years, the third in 4 years. The employment between 1981 and 1985 was a shock reaction caused by the banking system's collective action of substantially increasing the interest rates in a phase of deep stagnation. The dashed curve shows the envelope for oscillations after the first shock with this theory's result for the system's natural damping constant of 0.5 p.a. Correcting the data with this damping for all four shock reactions would bring them inside the range of the ellipse.

The first recording of cycles in social systems is due to the Hudson Bay Company. Its bookkeepers discovered the correlation between the trapper's fur deliveries of rabbits and foxes. Ginzburg and Colyvan gave a general survey of predator-prey cycles [5]. Without realizing its broader use a comprehensive study was made with a Nd:YAG laser using exited states as prey and cavity photons as predators [6]. The complete range between small signal oscillations and full "market" clearance with sharp periodic spikes could be obtained with constant supply of exited states and very small modulations of the cavity loss. Here we study business cycles excited with shocks or small modulations of the supply of consumer goods and compare the result with the data of figure 1.

The main physical problems with understanding business cycles were so far the lack of a resonator and of a mechanism fixing the observed phase shift between employment and inflation. We remove these barriers by dividing the process into the sequential subsystems production, employment, consumption, annual output, money supply, and inflation. The consumer market is simultaneously stimulated by supply and demand. When $p(t)$ goods and services are produced within a constant time horizon $\tau_p$ and $q(t)$ are consumed within $\tau_q$ the annual flows are given by

$$\dot{q} = pq/\bar{p}\tau_q - q/\tau_q \tag{1}$$

for consumers and

$$\dot{p} = \bar{p}/\tau_p - pq/\bar{p}\tau_q \tag{2}$$

for producers. The market term's denominator secures equilibrium $\bar{p}/\tau_p = \bar{q}/\tau_q$ for stable flows. An analytic solution can be obtained for small oscillations of $p = \bar{p}(1+u)$ and $q = \bar{q}(1+v)$. Then the product $\hat{u}\hat{v} \ll \hat{u}, \hat{v} \ll 1$ of the amplitudes can be neglected. (1) yields

$$\dot{v} = u/\tau_p \tag{3}$$

and fixes the phase delay $\psi = \pi/2$ of $v$ with respect to $u$. Eliminating the $p$-dependence of (2) with (3) and is time derivative results in the harmonic oscillator equation

$$\ddot{v} + \dot{v}/\tau_p + v/\tau_p\tau_q = 0 \tag{4}$$

with damping constant $d = 1/2\tau_p$ and base period $T_o = 2\pi/\omega_o = 2\pi(\tau_p\tau_q)^{1/2}$. Without a good reason for different time horizons the observed cycle period of 7 years means that the calendar year $\tau_o$ is still the dominant time horizon for producers and consumers. With $\tau_p = \tau_q = \tau_o$ the theoretical cycle period becomes $T_c = 2\pi/(\omega_o^2 - d^2)^{1/2} = 4\pi\tau_o/3^{1/2} = 7.3$ years.

The dashed envelope in figure 1 shows the decay $\exp(-t/2\tau_o)$ of the amplitude caused by the first oil price shock. After $t = T_c$ only 3 % propagate into the following cycle. This confirms the view that spontaneous cycles cannot exist. But there is a second solution for collective anticipation. Extending (2) with a modulation $\bar{p}\hat{r}e^{i\omega t}$ of production leads to a term $\omega_o^2 \hat{r} e^{i\omega t}$ on the right hand side of (4). Substituting $v_\omega = \hat{v}_\omega e^{i\omega t}$ for $v$ in (4) yields

$$v_\omega(1 - \omega^2/\omega_o^2 + i\omega/\omega_o) = r_\omega \quad . \tag{5}$$

Resonant modulation $\omega = \omega_o$ yields with $v(\omega_o, t) = -i\, r(\omega_o, t)$ the optimal phase shift $\pi/2$ for cycles with no damping at all. This explains the system's sensitivity to collective behavior. The pattern observed with figure 1 is a superposition of shock reaction and anticipation.

The ellipse's component
$$e = e_o - \hat{e} \cos \omega t \qquad (6)$$
for employment is in phase with production. It is plotted in figure 1 with amplitude $\hat{e} = 2\%$ and centered at the medium-term average $e_o = 94\%$ of full employment. Since consumption is delayed by $\psi = \pi/2$ with respect to production and since they stimulate the consumer market together the annual output is stabilized at a phase delay of $\varphi = \pi/4$ or T/8 with respect to production. The same delay is expected for the oscillation of the entire national output and the resulting oscillation $m$ of the average money supply $\bar{m}$.

Special drivers of inflation are price increases of imports and public debt. The difference between the money supply rate and the real structural growth rate is the general driver of inflation. This fixes the observed phase delay $\varphi = \pi/4$ of inflation with respect to employment and production.

The increasing semi-cycle of the ellipse is amplified for 2 to 3 years by positive feedback. But its accumulated money supply is generally not removed in the decreasing phase. For strong decreases governments and/or Central Banks inject even additional money hoping structural growth will justify their intervention later. Since this could not happen for the small structural growth rate of USA the 32 annual rates accumulated to a factor 4.62 for the currency's inflation in 32 years. The mean annual inflation rate was 4.9 % p.a.

The inflation component
$$j = j_o - \hat{j} \cos(\omega t - \pi/4) \qquad (7)$$
is centered at a medium-term level of $j_o = 4.0\%$ p. a. The oscillating amplitude of $\hat{j} = 3\%$ p.a. suggests an average collective amplification factor of $\hat{j}/\hat{e} = 1.5$ in the increasing semi-cycle. This factor and the delay $\varphi = \pi/4$ fixed the ellipse's main axes and their direction in figure 1.

The difference of 0.9 % p.a. between ellipse's and the data's mean inflation may be due to structural inflation. In a parallel paper we derive analytic solutions for structural growth. Combining both papers allows separating speculative from actually required investment and business cycles from structural growth. In this long-term perspective the best protection from inflation is not to intervene but to rely on the high natural damping constant of $1/2\tau_o$.